\documentclass[12pt,letterpaper]{article}

\usepackage{arxiv}
\usepackage{tikz}
\newcommand{\pidffmodel}{
    \begin{tikzpicture}[x=0.75pt,y=0.75pt,yscale=-1,xscale=1]
    
    \draw  [fill={rgb, 255:red, 255; green, 254; blue, 156 }  ,fill opacity=1 ] (278.88,197.09) -- (366.89,197.09) -- (366.89,244.55) -- (278.88,244.55) -- cycle ;
    \draw  [fill={rgb, 255:red, 255; green, 156; blue, 156 }  ,fill opacity=1 ] (278.88,254.04) -- (366.89,254.04) -- (366.89,301.49) -- (278.88,301.49) -- cycle ;
    \draw  [fill={rgb, 255:red, 200; green, 255; blue, 156 }  ,fill opacity=1 ] (278.88,141) -- (366.89,141) -- (366.89,188.46) -- (278.88,188.46) -- cycle ;
    \draw  [fill={rgb, 255:red, 197; green, 156; blue, 255 }  ,fill opacity=1 ] (277.88,82.71) -- (365.89,82.71) -- (365.89,130.17) -- (277.88,130.17) -- cycle ;
    \draw    (231.85,166.02) -- (277.74,166.02) ;
    \draw [shift={(279.74,166.02)}, rotate = 180] [color={rgb, 255:red, 0; green, 0; blue, 0 }  ][line width=0.75]    (10.93,-3.29) .. controls (6.95,-1.4) and (3.31,-0.3) .. (0,0) .. controls (3.31,0.3) and (6.95,1.4) .. (10.93,3.29)   ;
    \draw    (190.14,221.25) -- (277.74,221.25) ;
    \draw [shift={(279.74,221.25)}, rotate = 180] [color={rgb, 255:red, 0; green, 0; blue, 0 }  ][line width=0.75]    (10.93,-3.29) .. controls (6.95,-1.4) and (3.31,-0.3) .. (0,0) .. controls (3.31,0.3) and (6.95,1.4) .. (10.93,3.29)   ;
    \draw    (232.28,278.2) -- (276.88,278.2) ;
    \draw [shift={(278.88,278.2)}, rotate = 180] [color={rgb, 255:red, 0; green, 0; blue, 0 }  ][line width=0.75]    (10.93,-3.29) .. controls (6.95,-1.4) and (3.31,-0.3) .. (0,0) .. controls (3.31,0.3) and (6.95,1.4) .. (10.93,3.29)   ;
    \draw  [fill={rgb, 255:red, 161; green, 232; blue, 253 }  ,fill opacity=1 ] (147,220.72) .. controls (147,208.81) and (156.65,199.15) .. (168.57,199.15) .. controls (180.48,199.15) and (190.14,208.81) .. (190.14,220.72) .. controls (190.14,232.64) and (180.48,242.29) .. (168.57,242.29) .. controls (156.65,242.29) and (147,232.64) .. (147,220.72) -- cycle ;
    \draw  [fill={rgb, 255:red, 253; green, 198; blue, 161 }  ,fill opacity=1 ] (411.76,219.52) .. controls (411.76,207.61) and (421.42,197.95) .. (433.33,197.95) .. controls (445.25,197.95) and (454.91,207.61) .. (454.91,219.52) .. controls (454.91,231.44) and (445.25,241.09) .. (433.33,241.09) .. controls (421.42,241.09) and (411.76,231.44) .. (411.76,219.52) -- cycle ;
    \draw    (366.89,221.25) -- (409.76,221.25) ;
    \draw [shift={(411.76,221.25)}, rotate = 180] [color={rgb, 255:red, 0; green, 0; blue, 0 }  ][line width=0.75]    (10.93,-3.29) .. controls (6.95,-1.4) and (3.31,-0.3) .. (0,0) .. controls (3.31,0.3) and (6.95,1.4) .. (10.93,3.29)   ;
    \draw    (366.76,277.33) -- (382,277.33) ;
    \draw    (366.89,167.75) -- (382,167.75) ;
    \draw    (382,167.75) -- (414.84,200.59) ;
    \draw [shift={(416.25,202)}, rotate = 225] [color={rgb, 255:red, 0; green, 0; blue, 0 }  ][line width=0.75]    (10.93,-3.29) .. controls (6.95,-1.4) and (3.31,-0.3) .. (0,0) .. controls (3.31,0.3) and (6.95,1.4) .. (10.93,3.29)   ;
    \draw    (382,277.33) -- (416.65,239.48) ;
    \draw [shift={(418,238)}, rotate = 492.47] [color={rgb, 255:red, 0; green, 0; blue, 0 }  ][line width=0.75]    (10.93,-3.29) .. controls (6.95,-1.4) and (3.31,-0.3) .. (0,0) .. controls (3.31,0.3) and (6.95,1.4) .. (10.93,3.29)   ;
    \draw    (539.47,219.52) -- (539.47,256.35) ;
    \draw [shift={(539.47,258.35)}, rotate = 270] [color={rgb, 255:red, 0; green, 0; blue, 0 }  ][line width=0.75]    (10.93,-3.29) .. controls (6.95,-1.4) and (3.31,-0.3) .. (0,0) .. controls (3.31,0.3) and (6.95,1.4) .. (10.93,3.29)   ;
    \draw  [fill={rgb, 255:red, 182; green, 255; blue, 215 }  ,fill opacity=1 ] (512.72,258.35) -- (573.12,258.35) -- (573.12,292.87) -- (512.72,292.87) -- cycle ;
    \draw    (127.5,108) -- (275,108) ;
    \draw [shift={(277,108)}, rotate = 180] [color={rgb, 255:red, 0; green, 0; blue, 0 }  ][line width=0.75]    (10.93,-3.29) .. controls (6.95,-1.4) and (3.31,-0.3) .. (0,0) .. controls (3.31,0.3) and (6.95,1.4) .. (10.93,3.29)   ;
    \draw    (454.91,219.52) -- (539.47,219.52) ;
    \draw    (127.5,108) -- (127.5,220.36) ;
    \draw    (108,220) -- (145,220.69) ;
    \draw [shift={(147,220.72)}, rotate = 181.06] [color={rgb, 255:red, 0; green, 0; blue, 0 }  ][line width=0.75]    (10.93,-3.29) .. controls (6.95,-1.4) and (3.31,-0.3) .. (0,0) .. controls (3.31,0.3) and (6.95,1.4) .. (10.93,3.29)   ;
    \draw    (574,277) -- (600,277) ;
    \draw    (600,277) -- (600.5,318) ;
    \draw    (168.07,318) -- (600.5,318) ;

    \draw    (168.07,318) -- (168.56,244.29) ;
    \draw [shift={(168.57,242.29)}, rotate = 450.38] [color={rgb, 255:red, 0; green, 0; blue, 0 }  ][line width=0.75]    (10.93,-3.29) .. controls (6.95,-1.4) and (3.31,-0.3) .. (0,0) .. controls (3.31,0.3) and (6.95,1.4) .. (10.93,3.29)   ;
    \draw    (231.85,166.02) -- (231.85,278.2) ;
    \draw    (367,106) -- (434,106) ;
    \draw    (433.33,106) -- (433.33,195.95) ;
    \draw [shift={(433.33,197.95)}, rotate = 270] [color={rgb, 255:red, 0; green, 0; blue, 0 }  ][line width=0.75]    (10.93,-3.29) .. controls (6.95,-1.4) and (3.31,-0.3) .. (0,0) .. controls (3.31,0.3) and (6.95,1.4) .. (10.93,3.29)   ;
    
    \draw (296.16,156.3) node [anchor=north west][inner sep=0.75pt]  [font=\Large] [align=left] {$\displaystyle K_{p}$$\displaystyle e( t)$};
    \draw (283.68,204.38) node [anchor=north west][inner sep=0.75pt]  [font=\small] [align=left] {$\displaystyle K_{i}$$\displaystyle \sum ^{k}_{i=1} e( t_{i} ) \Delta t $};
    \draw (277.16,267.2) node [anchor=north west][inner sep=0.75pt]  [font=\scriptsize] [align=left] {$\displaystyle K_{d}$$\displaystyle \frac{e(t_{k})-e(t_{k+1})}{\Delta t}$};
    \draw (171.27,200.51) node [anchor=north west][inner sep=0.75pt]   [align=left] {$ $};
    \draw (159.57,211.0) node [anchor=north west][inner sep=0.75pt]  [font=\LARGE]  {${\displaystyle \Sigma}$};
    \draw (200.58,200.6) node [anchor=north west][inner sep=0.75pt]   [align=left] {$\displaystyle e( t)$};
    \draw (79.08,211.45) node [anchor=north west][inner sep=0.75pt]   [align=left] {$\displaystyle r( t)$};
    \draw (306.09,97.74) node [anchor=north west][inner sep=0.75pt]  [font=\Large] [align=left] {$\displaystyle FF$};
    \draw (424.33,210.99) node [anchor=north west][inner sep=0.75pt]  [font=\LARGE]  {${\displaystyle \Sigma}$};
    \draw (527.06,264.75) node [anchor=north west][inner sep=0.75pt]  [font=\Large] [align=left] {$\displaystyle u( t)$};
    \draw (373.04,83.45) node [anchor=north west][inner sep=0.75pt]   [align=left] {$\displaystyle baseline$};
    \draw (461.12,198.96) node [anchor=north west][inner sep=0.75pt]   [align=left] {$\displaystyle corrective$};
    \draw (184,240.4) node [anchor=north west][inner sep=0.75pt]    {$-$};
    \draw (129.5,227.76) node [anchor=north west][inner sep=0.75pt]    {$+$};
    \end{tikzpicture}
}

\newcommand{\pidffmp}{
\begin{tikzpicture}[x=0.75pt,y=0.75pt,yscale=-1,xscale=1]
    
    \draw  [fill={rgb, 255:red, 255; green, 254; blue, 156 }  ,fill opacity=1 ] (279.88,310.09) -- (367.89,310.09) -- (367.89,357.55) -- (279.88,357.55) -- cycle ;
    \draw  [fill={rgb, 255:red, 255; green, 156; blue, 156 }  ,fill opacity=1 ] (279.88,367.04) -- (367.89,367.04) -- (367.89,414.49) -- (279.88,414.49) -- cycle ;
    \draw  [fill={rgb, 255:red, 200; green, 255; blue, 156 }  ,fill opacity=1 ] (279.88,254) -- (367.89,254) -- (367.89,301.46) -- (279.88,301.46) -- cycle ;
    \draw  [fill={rgb, 255:red, 197; green, 156; blue, 255 }  ,fill opacity=1 ] (279.88,139.71) -- (367.89,139.71) -- (367.89,187.17) -- (279.88,187.17) -- cycle ;
    \draw    (232.85,279.02) -- (278.74,279.02) ;
    \draw [shift={(280.74,279.02)}, rotate = 180] [color={rgb, 255:red, 0; green, 0; blue, 0 }  ][line width=0.75]    (10.93,-3.29) .. controls (6.95,-1.4) and (3.31,-0.3) .. (0,0) .. controls (3.31,0.3) and (6.95,1.4) .. (10.93,3.29)   ;
    \draw    (191.14,334.25) -- (278.74,334.25) ;
    \draw [shift={(280.74,334.25)}, rotate = 180] [color={rgb, 255:red, 0; green, 0; blue, 0 }  ][line width=0.75]    (10.93,-3.29) .. controls (6.95,-1.4) and (3.31,-0.3) .. (0,0) .. controls (3.31,0.3) and (6.95,1.4) .. (10.93,3.29)   ;
    \draw    (233.28,391.2) -- (277.88,391.2) ;
    \draw [shift={(279.88,391.2)}, rotate = 180] [color={rgb, 255:red, 0; green, 0; blue, 0 }  ][line width=0.75]    (10.93,-3.29) .. controls (6.95,-1.4) and (3.31,-0.3) .. (0,0) .. controls (3.31,0.3) and (6.95,1.4) .. (10.93,3.29)   ;
    \draw  [fill={rgb, 255:red, 161; green, 232; blue, 253 }  ,fill opacity=1 ] (148,333.72) .. controls (148,321.81) and (157.65,312.15) .. (169.57,312.15) .. controls (181.48,312.15) and (191.14,321.81) .. (191.14,333.72) .. controls (191.14,345.64) and (181.48,355.29) .. (169.57,355.29) .. controls (157.65,355.29) and (148,345.64) .. (148,333.72) -- cycle ;
    \draw  [fill={rgb, 255:red, 253; green, 198; blue, 161 }  ,fill opacity=1 ] (412.76,332.52) .. controls (412.76,320.61) and (422.42,310.95) .. (434.33,310.95) .. controls (446.25,310.95) and (455.91,320.61) .. (455.91,332.52) .. controls (455.91,344.44) and (446.25,354.09) .. (434.33,354.09) .. controls (422.42,354.09) and (412.76,344.44) .. (412.76,332.52) -- cycle ;
    \draw    (367.89,334.25) -- (410.76,334.25) ;
    \draw [shift={(412.76,334.25)}, rotate = 180] [color={rgb, 255:red, 0; green, 0; blue, 0 }  ][line width=0.75]    (10.93,-3.29) .. controls (6.95,-1.4) and (3.31,-0.3) .. (0,0) .. controls (3.31,0.3) and (6.95,1.4) .. (10.93,3.29)   ;
    \draw    (367.76,390.33) -- (383,390.33) ;
    \draw    (367.89,280.75) -- (383,280.75) ;
    \draw    (383,280.75) -- (417.59,315.58) ;
    \draw [shift={(419,317)}, rotate = 225.2] [color={rgb, 255:red, 0; green, 0; blue, 0 }  ][line width=0.75]    (10.93,-3.29) .. controls (6.95,-1.4) and (3.31,-0.3) .. (0,0) .. controls (3.31,0.3) and (6.95,1.4) .. (10.93,3.29)   ;
    \draw    (383,390.33) -- (418.65,351.47) ;
    \draw [shift={(420,350)}, rotate = 492.53] [color={rgb, 255:red, 0; green, 0; blue, 0 }  ][line width=0.75]    (10.93,-3.29) .. controls (6.95,-1.4) and (3.31,-0.3) .. (0,0) .. controls (3.31,0.3) and (6.95,1.4) .. (10.93,3.29)   ;
    \draw    (540.47,332.52) -- (540.47,369.35) ;
    \draw [shift={(540.47,371.35)}, rotate = 270] [color={rgb, 255:red, 0; green, 0; blue, 0 }  ][line width=0.75]    (10.93,-3.29) .. controls (6.95,-1.4) and (3.31,-0.3) .. (0,0) .. controls (3.31,0.3) and (6.95,1.4) .. (10.93,3.29)   ;
    \draw  [fill={rgb, 255:red, 182; green, 255; blue, 215 }  ,fill opacity=1 ] (513.72,371.35) -- (574.12,371.35) -- (574.12,405.87) -- (513.72,405.87) -- cycle ;
    \draw    (129.5,165) -- (277,165) ;
    \draw [shift={(279,165)}, rotate = 180] [color={rgb, 255:red, 0; green, 0; blue, 0 }  ][line width=0.75]    (10.93,-3.29) .. controls (6.95,-1.4) and (3.31,-0.3) .. (0,0) .. controls (3.31,0.3) and (6.95,1.4) .. (10.93,3.29)   ;
    \draw    (455.91,332.52) -- (540.47,332.52) ;
    \draw    (128.5,165) -- (128.5,333.36) ;
    \draw    (109,333) -- (146,333.69) ;
    \draw [shift={(148,333.72)}, rotate = 181.06] [color={rgb, 255:red, 0; green, 0; blue, 0 }  ][line width=0.75]    (10.93,-3.29) .. controls (6.95,-1.4) and (3.31,-0.3) .. (0,0) .. controls (3.31,0.3) and (6.95,1.4) .. (10.93,3.29)   ;
    \draw    (575,390) -- (601,390) ;
    \draw    (601,390) -- (601.5,431) ;
    \draw    (169.07,431) -- (601.5,431) ;
    \draw    (169.07,431) -- (169.56,357.29) ;
    \draw [shift={(169.57,355.29)}, rotate = 450.38] [color={rgb, 255:red, 0; green, 0; blue, 0 }  ][line width=0.75]    (10.93,-3.29) .. controls (6.95,-1.4) and (3.31,-0.3) .. (0,0) .. controls (3.31,0.3) and (6.95,1.4) .. (10.93,3.29)   ;
    \draw    (232.85,279.02) -- (232.85,391.2) ;
    \draw    (369,163) -- (491,163) ;
    \draw    (491,279) -- (452.78,315.57) ;
    \draw [shift={(451.33,316.95)}, rotate = 316.27] [color={rgb, 255:red, 0; green, 0; blue, 0 }  ][line width=0.75]    (10.93,-3.29) .. controls (6.95,-1.4) and (3.31,-0.3) .. (0,0) .. controls (3.31,0.3) and (6.95,1.4) .. (10.93,3.29)   ;
    \draw    (213,224) -- (213,334) ;
    \draw    (213,224) -- (277,224) ;
    \draw [shift={(279,224)}, rotate = 180] [color={rgb, 255:red, 0; green, 0; blue, 0 }  ][line width=0.75]    (10.93,-3.29) .. controls (6.95,-1.4) and (3.31,-0.3) .. (0,0) .. controls (3.31,0.3) and (6.95,1.4) .. (10.93,3.29)   ;
    \draw  [fill={rgb, 255:red, 243; green, 131; blue, 98 }  ,fill opacity=1 ] (279.88,198.71) -- (367.89,198.71) -- (367.89,246.17) -- (279.88,246.17) -- cycle ;
    \draw    (434.33,224) -- (434.33,308.95) ;
    \draw [shift={(434.33,310.95)}, rotate = 270] [color={rgb, 255:red, 0; green, 0; blue, 0 }  ][line width=0.75]    (10.93,-3.29) .. controls (6.95,-1.4) and (3.31,-0.3) .. (0,0) .. controls (3.31,0.3) and (6.95,1.4) .. (10.93,3.29)   ;
    \draw    (368,224) -- (434.33,224) ;
    \draw    (491,162) -- (491,279) ;
    
    \draw (300.16,268.3) node [anchor=north west][inner sep=0.75pt]  [font=\Large] [align=left] {$\displaystyle K_{p}$$\displaystyle e( t)$};
     \draw (283.68,314.38) node [anchor=north west][inner sep=0.75pt]  [font=\small] [align=left] {$\displaystyle K_{i}$$\displaystyle \sum ^{k}_{i=1} e( t_{i} ) \Delta t $};
    \draw (278.16,378) node [anchor=north west][inner sep=0.75pt]  [font=\scriptsize] [align=left] {$\displaystyle K_{d}$$\displaystyle \frac{e(t_{k})-e(t_{k+1})}{\Delta t}$};
    \draw (172.27,313.51) node [anchor=north west][inner sep=0.75pt]   [align=left] {$ $};
    \draw (159.57,324.19) node [anchor=north west][inner sep=0.75pt]  [font=\LARGE]  {${\displaystyle \Sigma}$};
    \draw (199.58,335.6) node [anchor=north west][inner sep=0.75pt]   [align=left] {$\displaystyle e( t)$};
    \draw (80.08,324.45) node [anchor=north west][inner sep=0.75pt]   [align=left] {$\displaystyle r( t)$};
    \draw (308.09,155.74) node [anchor=north west][inner sep=0.75pt]  [font=\Large] [align=left] {$\displaystyle FF$};
    \draw (425.33,324.99) node [anchor=north west][inner sep=0.75pt]  [font=\LARGE]  {${\displaystyle \Sigma}$};
    \draw (528.06,377.75) node [anchor=north west][inner sep=0.75pt]  [font=\Large] [align=left] {$\displaystyle u( t)$};
    \draw (400.04,142.45) node [anchor=north west][inner sep=0.75pt]   [align=left] {$\displaystyle baseline$};
    \draw (462.12,311.96) node [anchor=north west][inner sep=0.75pt]   [align=left] {$\displaystyle corrective$};
    \draw (185,353.4) node [anchor=north west][inner sep=0.75pt]    {$-$};
    \draw (130.5,340.76) node [anchor=north west][inner sep=0.75pt]    {$+$};
    \draw (289.09,215.74) node [anchor=north west][inner sep=0.75pt]  [font=\normalsize] [align=left] {$\displaystyle K_{m} M( e( t))$};
    \draw (377.89,200.71) node [anchor=north west][inner sep=0.75pt]   [align=left] {$\displaystyle predictive$};
        \end{tikzpicture}

}

\usepackage{cite}
\usepackage[utf8]{inputenc} 
\usepackage[T1]{fontenc}    
\usepackage{hyperref}       
\usepackage{url}            
\usepackage{booktabs}       
\usepackage{amsfonts}       
\usepackage{nicefrac}       
\usepackage{microtype}      
\usepackage{lipsum}
\usepackage{graphicx}
\usepackage{setspace}

\usepackage[font=small,labelfont=bf]{caption} 
\usepackage{tikz}
\usetikzlibrary{shapes,arrows}
\graphicspath{ {./images/} }
\tikzset{every picture/.style={line width=0.755pt}} 

\title{Cyclical Electromechanical Error Denial System Using Matrix Profile }

\author{
  Peter Bowman-Davis \\
  Justin-Siena High School\\
  Napa, CA 94558 \\
  \texttt{peterbowmandavis21@js-student.org} \\
   \And
  Gary Laski \\
  Georgia Institute of Technology\\
  Atlanta, GA 30332 \\
  \texttt{glaski3@gatech.edu} \\
}

\begin{document}
\setlength{\headheight}{15pt}
\setlength{\parskip}{0pt}
\maketitle

\begin{abstract}
We propose that the Matrix Profile data structure, conventionally applied to large scale time-series data mining, is applicable to the analysis and suppression of cyclical error in electromechanical systems, paving the way for an intelligent family of adaptable control systems which respond to environmental error at a computational cost low enough to be practical in embedded applications. We construct and evaluate the efficacy of a control algorithm utilizing the Matrix Profile, which we call Cyclical Electromechanical Error Denial System (CEEDS).
\end{abstract}

\section{Introduction}
\begin{figure}[h]
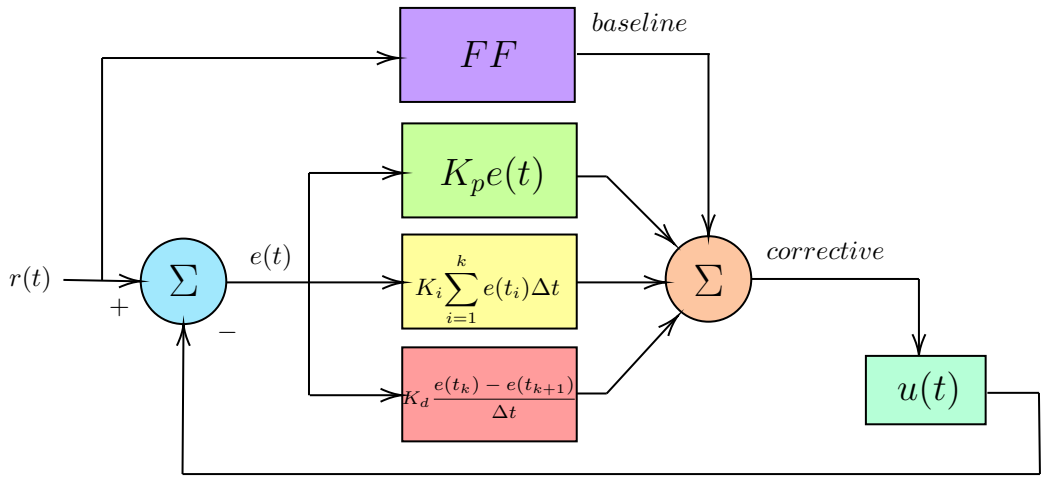

    \centering
    \pidffmodel
    
    \caption{Model of a standard PID control system with Feed Forward}
    \label{fig:pidffmodel}
\end{figure}

The Proportional-Integral-Derivative algorithm (PID), pioneered over a hundred years ago, has proven to be one of the most influential algorithms in the field of control theory. It has been implemented in a multitude of control systems varying greatly in scale, function, and importance. A nearly universally applicable but less commonly used improvement upon PID systems involves the generation and application of a transfer function, which maps control inputs to metric-based outputs.
From this transfer function, inputting the setpoint into the inverse will return a control signal that is fed forward into the system to move it toward the desired state. The PID control can then use feedback to correct for any small environmental errors, rather than the whole state of the system. This algorithm forms the base on which CEEDS expands using the Matrix Profile data structure, and can be described by the following equation : 
\begin{equation}
\label{eqn:control_function}
  u(t) = K_p e(t)+K_i\int_{0}^{t} e(t') dt' + K_d  \frac{de'(t)}{dt}
\end{equation}
While PID has incredible generality in application, it also does not go further than simple feedback control, and thus lacks the ability to predict error beyond the derivative term, which only looks at short term changes in the time series. In the field of controls, this has lead to the further development of related systems and attempted improvements ~\cite{pidIntegrating, LimitsPID1,LimitsPID2} in the direction of applying PID-based controllers to higher complexity signal denial systems. Another approach to enhancing this control system is the use of machine learning algorithms to analyze error in control systems ~\cite{nnPID} predicatively rather than correctively. However, such systems are not necessarily as stable as compared to their hard coded counterparts ~\cite{nnStability}, as these types of algorithms are black-box, and do not offer insight as to how they actually work, other than providing a heuristic equation to suppress error signals. Additionally, they are computationally expensive ~\cite{nnSUCKS},  which makes machine learning a sub-optimal choice for real-time control on embedded systems. As a result, it can be difficult to assume that a machine learning based control system will respond effectively to forms of error which the algorithm hasn't been trained or tested on. 
Therefore, a new type of control system is necessary for control plants which may encounter such cyclical error: one which is able to improve upon the PID algorithm and remain computationally cheap enough for usage in a wide variety of applications.
\break \break
In this paper, we propose the usage of the Matrix Profile, a mathematical structure typically applied to data mining of large scale time series data sets, to detect and proactively correct for recurrent error patterns, a common issue in many control plants ~\cite{telescoper}, particularly those which deal with rotational motion. Computing the Matrix Profile allows for periodic motif detection, which then can be implemented in  controls, allowing the electromechanical system to intuitively correct for periodic error patterns. While such a methodology is not the typical use case for the Matrix Profile, it is well suited to this task due to its low computational footprint \cite{mpSUPERHUMAN} as well as its ability to generalize error patterns \cite{mpMOTIFS}. We propose the initial implementation of this algorithm in our electromechanical system under the pretense of a cyclical error, as such errors are both present in the real world, and serve as a good test case for the feasibility of the Matrix Profile data structure in sub 1 GHz machines. As an addendum, the preceding PID Feed Forward system can be described in a control model by Figure \ref{fig:pidffmodel}. Note that the PIDF equation is in continuous time whereas the functional model describes a discrete time series.

\section{Implementation}
\label{sec:headings}
In order to test our application, we constructed two physical platforms to generate data that will allow us to determine the practicality CEEDS in a real-world environment. By testing upon physical platforms, we are able to observe not only the theoretical success of CEEDS, but also its ability to perform in an environment which has elements such as stochastic noise in data and discrete error geometries. To perform the collection of such data, we selected the following commercially available components for their low cost and simplicity of operation.

\paragraph{Physical Test Beds.}
Our application was tested on two identical physical setups, containing 12V W39530 TETRIX® MAX DC Motors and E4T Rotary Encoders controlled by a PWM signal through a IRFZ44 N-Channel MOSFET for each platform. Both setups utilize a 700 MHz ARM1176JZF-S (1 Core 1 Thread) processor aboard a Raspberry Pi Generation 1 Model B, which we use to program and deploy CEEDS through the usage of the GPIO pins of the Raspberry Pi. 

\paragraph{CEEDS Implementation.} 
Both Raspberry Pi Model B boards run the Raspbian Buster Lite Headless distribution. CEEDS is currently written in Python 2.7 using the Matrix Profile Foundation's MatrixProfile library ~\cite{VanBenschoten2020} as the primary dependency. In order to generate test data with cyclical error, we impose a software based interference signal that is able to be modified in frequency, amplitude, and shape. In this way, we can test the robustness of CEEDS in a variety of situations. Such an interference could alternatively be applied mechanically, but this would complicate the physical test setup and limit the range of error signals we apply.

\section{Methodology}
\label{sec:implementaion}
\paragraph{Data Collection.} 
CEEDS commences by running the motor to a setpoint RPM inputted by the user. This is achieved through the use of the transfer function to translate the desired RPM to a PWM duty cycle in hertz, correcting for any residual error through the use of a pre-tuned PID. It should be noted that our program is constructed inside of a loop which has a minimum runtime. This set runtime allows a host of advantages for the program, such as the accurate computation of the delta time variable in the PID system. Further, it allows for the translation of a data sample index to an actual point in the runtime. For example, because data is sampled every $t$ milliseconds, data index $i$ represents the datum sampled at time $t \cdot s$ where $s$ is the main loop runtime of CEEDS. In all of our example data, a runtime of 50 milliseconds was used, as it proved sufficient to produce loop iterations which rarely, if ever, exceeded 50 milliseconds. This intuitive method of data indexing allows CEEDS to accurately compute features of the cyclical error pattern, such as period and offset. This initial data collection phase is run until a user specified data sample cutoff is reached. 
\paragraph{Detection.}
The Matrix Profile data structure allows for the rapid analysis of large time series data and the discovery of "motifs" in the data – recurring shapes of data within the larger data set. In our application, we use such motifs as a classifier for repetitive error patterns.  After the set cutoff of run time is eclipsed, the Matrix Profile is computed using the builtin MPX algorithm from the "matrixprofile" library with the collected error data as an input. We chose to use this algorithm for computing the Matrix Profile as it is not only exact, but also extremely efficient, offering advantages in repeatability and performance in our system compared to algorithms such as STAMP ~\cite{algorithms}. Before such an algorithm is executed however, CEEDS sets the motor's RPM to the setpoint RPM generated by the transfer function in order to mitigate anomalies in PID feedback created by runtime holds in the thread. It should be noted that in a multithreaded system, the computation of the Matrix Profile could be carried out in a separate worker thread, allowing the PID element to be retained in the primary control thread.

\paragraph{Analysis.}
After the Matrix Profile is computed from the data, CEEDS computes the top $n$ motifs from the data, where $n$ is a constant predefined by the user. In all of the data generated below, $n=5$. After testing various values of $n$, we found $n=5$ sufficiently diverse, but also a low enough value to keep computation time down in the next stage of CEEDS. A feature set is defined for each motif containing two distances: first, the "offset" of the interference from the start of the data, and second, the modal period between each successive motif. The modal aspect of such a feature set calculation allows for an accurate feature set even in the event of the motif finding sub-algorithm "skipping" a motif repetition.

\paragraph{Ranking.}
With such a feature set for each motif in the top $n$, CEEDS next flips the motif about the x-axis creating a cancellation signal. A list of zeroes of length $\ell$ is concatenated to the end of the cancellation signal, where $\ell$ represents the modal period minus the motif length. These cancellation signals are aligned according to their generated offset and periodicity, and duplicated across a list until the length of the cancellation list equals or is greater than the original data sample length. This cancellation signal is then truncated to the length of the original data sample's length. After this, each candidate cancellation signal list is added element-wise to a copy of the original logged error list, creating a retroactive cancellation list. This list simply represents the error over time had the motif been applied at the generated locations and offsets, allowing for ranking based upon reduction of said error. Each absolute value of the element of the candidate lists are then summed and ranked. The lowest value of such a computation should yield the most fit motif, given that the most fit cancellation signal should minimize the area under the error curve. From this ranking, one candidate motif is deemed the most fit and is passed onto the next stage of CEEDS. The structure of the candidate motif with proceeding zeros is as follows at this point, \[[y_{i1},y_{i2},y_{i3},y_{i4}...y_{in},0,0,0,0...0]\] where $y_i$ denotes an element from the generated motif. The buffer zeros after the $y$ terms represent the scaling of the length of the motif to fulfill its entire period. Thus, the length of such a list represents the modal at which the interference signal will be applied. 

\paragraph{Application.} 
After the most fit cancellation waveform and its feature set are determined, CEEDS enters the next stage, in which the cancellation waveform is applied cyclically. This is done through the iteration over the motif and proceeding zeroes every run time, then in turn taking this motif sample (or zero) and running it through our transfer function, which allows the conversion of RPM to an applied duty cycle. This should, given that the cyclical error persists, cancel the error, and create a steady state of error only affected by the stochastic friction forces present in the motor. Therefore, a new element $M(e(t))$ is added to the original control equation of PIDF, resulting in the following equation and corresponding model, where $M(x)$ is the Matrix Profile-based counter-interference signal CEEDS introduces: \begin{equation}
\label{eqn:control_function_ff}
  u(t) = K_p e(t)+K_i\int_{0}^{t} e(t') dt' + K_d  \frac{de'(t)}{dt}+K_m M(e(t)) + FF
\end{equation}

\begin{figure}[hbt!]
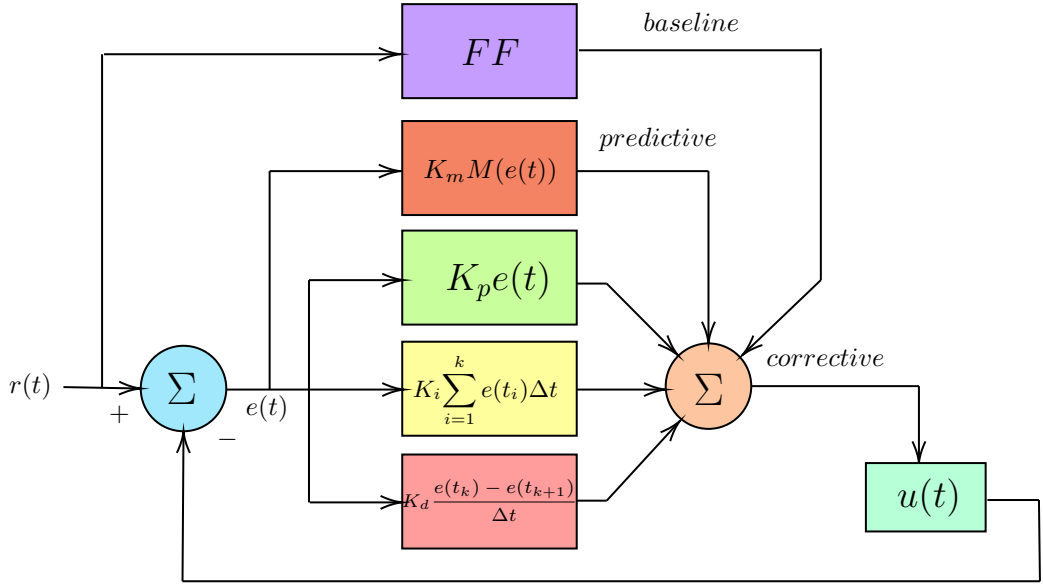

    \centering
    \pidffmp
    \caption{Model of our control system utilizing PID, Feed Forward, and Matrix Profile Analysis}
    \label{fig:pidffmp}
\end{figure}
It should be noted that alongside the addition of this new element, $M(e(t))$, a constant multiplier, $K_m$, is implemented to exert a greater control upon the effect of the MP element. While theoretically this element is optional, as a value of 1 should match the average profile of the interference signal, a higher or lower value may be dynamically changed in a more complex implementation. 

\section{Results}
After generating data on the physical test beds using the methodology outlined above, we next evaluate the efficacy of CEEDS by analyzing the resulting data. To verify the integrity of our results, multiple tests were run in each configuration and on each test setup. Repeated trials of each waveform can be located in the proceeding section, "Additional Figures", where at least one duplicate trial is provided for each waveform shape. The effectiveness of CEEDS can be determined by the comparison of the CEEDS-generated time series data to the PIDF data, which is generated in identical test conditions with the $K_m$ constant set to zero. For all CEEDS data, the novelty of CEEDS is only introduced after cutoff $y$ where $y$ is the data sample ID described by the ANALYSIS TIME = $y$*$n$ ms and $n$ represents the minimum number of milliseconds per runtime loop. This is due to the initial analysis window necessity of the PIDFM/CEEDS architecture. 
\break \break
It should be noted that as the current data is generated using a single threaded machine, sampling such data during the transition period between the data sample cutoff and the completion of the Matrix Profile computation is impossible as the thread is occupied with computing the Matrix Profile. To mitigate error in this section, the motor is set to the predicted power of the transfer function and the PID system is bypassed just before this computation takes place. Thus, any data in this region can be a described as the sum of the error of the transfer function's output with respect to the actual motor RPM, the introduced interference at this time, and finally, the stochastic noise in the system. In the future, a program running a non-standard sampling rate may allow for the integration of the PID system into this computation period, but this is not present in the current state of CEEDS.
\subsection{Square Waveform}

For the following figure, the initial system parameters were:
\\ WINDOW SIZE = 35 \\ ANALYSIS TIME = 600*50ms = 30 seconds

\begin{figure}[h!]
    \centering
    \captionsetup{type=figure}
    \includegraphics[scale=0.6]{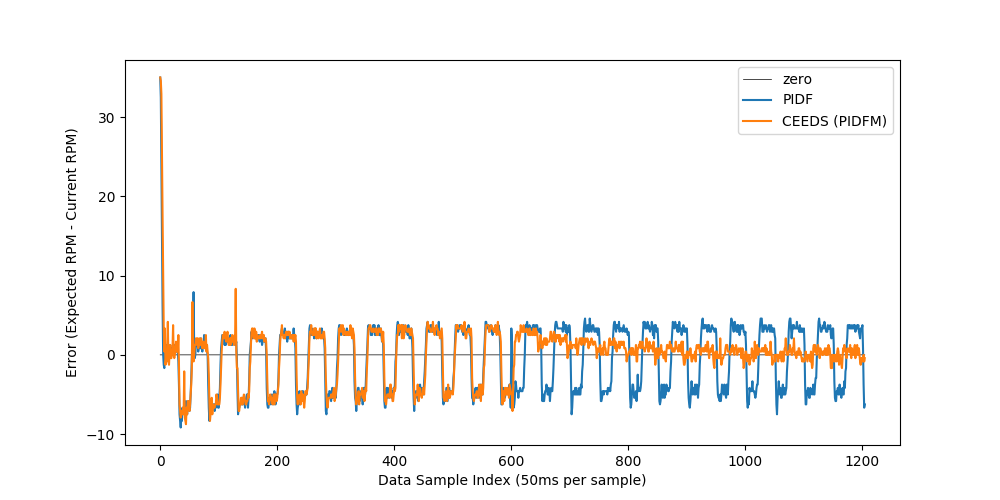}
    \captionof{figure}{Denial of square wave}
    \label{fig:square1}
\end{figure}

\begin{figure}[h!]
    \centering
    \captionsetup{type=figure}
    \includegraphics[scale=0.6]{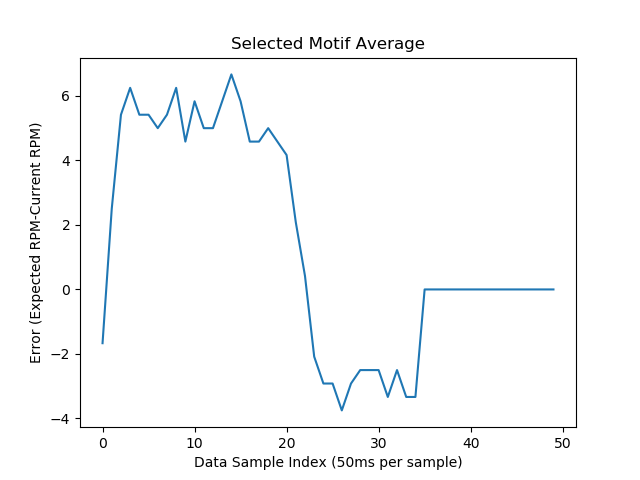}
    \captionof{figure}{Selected motif cancellation signal associated with Figure \ref{fig:square1} with zero padding}
    \label{fig:square2}
\end{figure}

Fig. \ref{fig:square1} shows the results of a square wave interference list of $[[0]*60+[50]*20]$ being denied, and Fig. \ref{fig:square2} shows the associated motif generated from the ranking system. It should be noted that Fig. \ref{fig:square2} as well as all the proceeding motif graphs are flipped about the x-axis to represent the average cancellation signal to be applied, and such a signal also contains the padding zeroes, which are in place in order to allow for the constant iteration through the signal while preserving the correct period of the cancellation signal. In this particular trial, a percentage reduction of the absolute value sum of error was calculated to be 40\% from data sample 55 until the end. Data sample 55 was used as a starting point due to the initial error of the motor start up sequence which stabilized in all samples by data sample 55. Also calculated was the reduction in error from data sample 600, when the Matrix Profile portion of CEEDS comes into effect. From this data sample to the end of the data, the absolute value sum of error was reduced by a staggering 73\%. It should be noted that the percentage reduction from sample 600 is a more accurate representation of the performance of the algorithm, as the reduction percentage from sample 50 will actually converge to the reduction percentage from sample 600 as runtime approaches infinity, providing the error remains cyclical and the period remains constant.

Fig. \ref{fig:square1} also demonstrates an unintended positive side effect of the matrix profile cancellation. The signal before CEEDS was placed into effect was never able to stabilize on zero despite the interference signal being only negative due to integral windup causing an over correction when the signal was not applied. By stabilizing and cancelling the signal, CEEDS provides the PIDF correction a signal which is more easily corrected by the integral term, bringing it to zero. This can be observed in Fig. \ref{fig:square1} in the data sample index region 601 to 820. This behavior is unexpected but beneficial to the stability of the system plant, due to the combined control obtained from both PIDF and Matrix Profile allowing for the complete stabilization of a signal.

\subsection{Negative Ramp Sawtooth Waveform}
\begin{figure}[h!]
    \centering
    \captionsetup{type=figure}
    \includegraphics[scale=0.6]{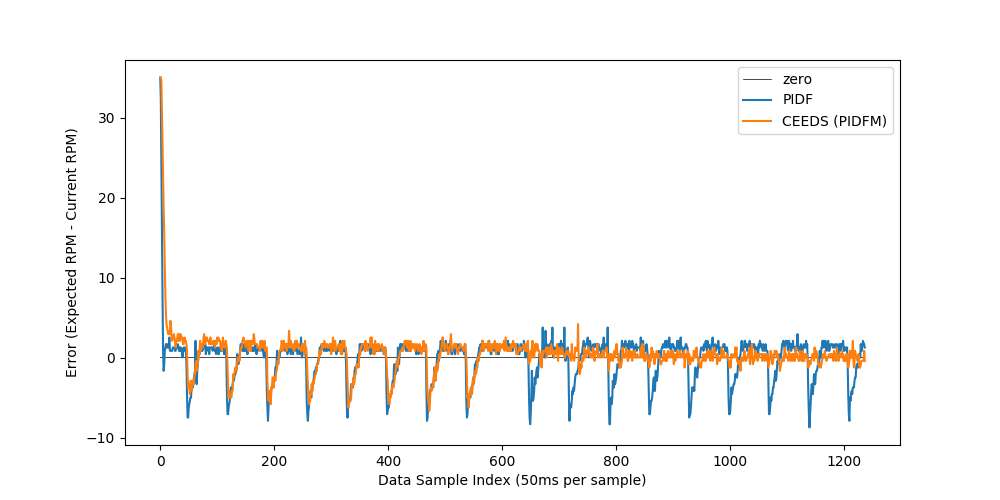}
    \captionof{figure}{Denial of negative ramp sawtooth wave}
    \label{fig:dec1}
\end{figure}

Fig. \ref{fig:dec1} demonstrates the denial of a triangular wave interference pattern of $[[0]*45+list(range(50,0,-2))]$.

In a second stage of testing, the system parameters were
\\ WINDOW SIZE = 35 \\ ANALYSIS TIME = 600*50ms = 30 seconds
\vspace{-4.2mm}

\newpage
It should be noted that such a wave is suppressed nearly completely by CEEDS, all while having the same initial parameters as the previous square signal. This in turn demonstrates CEEDS' ability to generalize waveform signals and accurately deny them thanks to flexibility of the Matrix Profile data structure. From data sample 55 to the end of the collected data, a 37\% error reduction was observed. From data sample 600 to the end of the collected data, a 71\% reduction in error was observed.
\vspace{-1mm}
\begin{figure}[h!]
    \centering
    \captionsetup{type=figure}
    \includegraphics[scale=0.6]{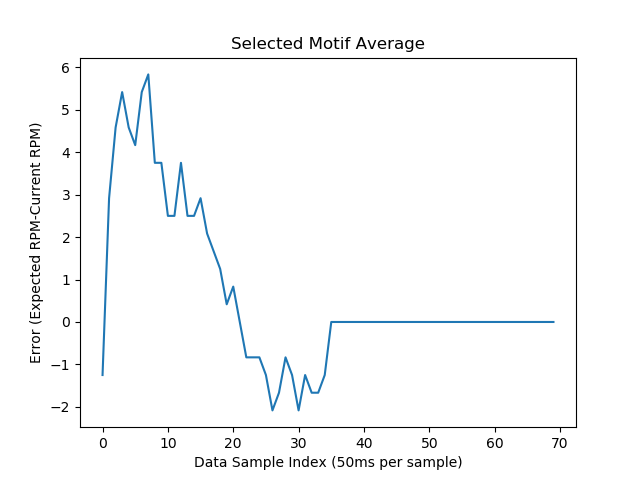}
    \captionof{figure}{Selected motif cancellation signal associated with Figure \ref{fig:dec1} with zero padding}
    \label{fig:dec2}
\end{figure}
\subsection{Triangle Waveform}
\begin{figure}[ht!]
    \centering
    \captionsetup{type=figure}
    \includegraphics[scale=0.6]{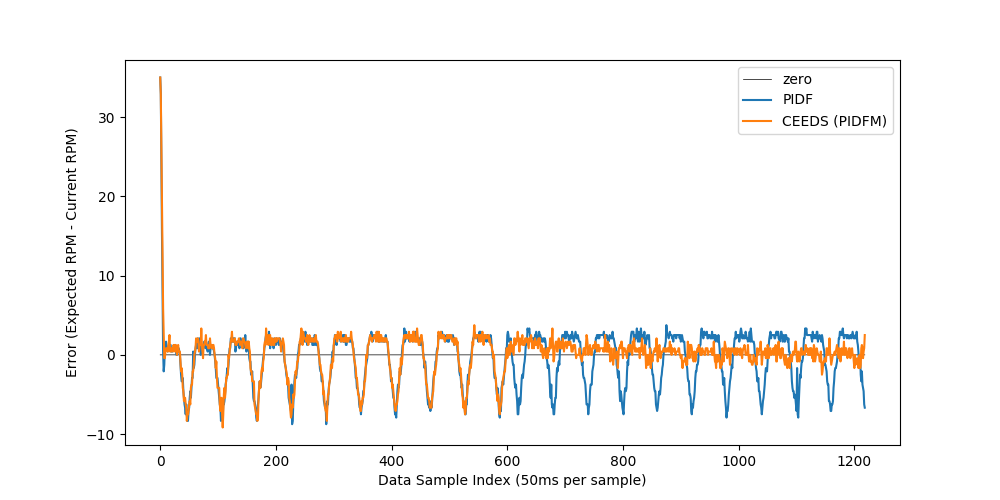}
    \captionof{figure}{Denial of triangular wave}
    \label{fig:tri1}
\end{figure}

Fig. \ref{fig:tri1} shows the results of a increasing-decreasing wave interference pattern of $[[0]*30+list(range(0,60,4))+list(range(60,0,-4))]$ in both applications.

As in the previous trials, the system parameters were as follows:
\\ WINDOW SIZE = 35 \\ ANALYSIS TIME = 600*50ms = 30 seconds
\vspace{-3.7mm}
\newpage
A 32\% reduction in absolute value error was calculated from data sample 55 to the end of the data, and a 72\% reduction in absolute value error was calculated from data sample 600 to the end of the data. The same signal convergence to zero seen in Fig. \ref{fig:square1} can also be observed in Fig. \ref{fig:tri1}, demonstrating the commonality of this behavior in a CEEDS integration. 
\begin{figure}[h!]
    \centering
    \captionsetup{type=figure}
    \includegraphics[scale=0.6]{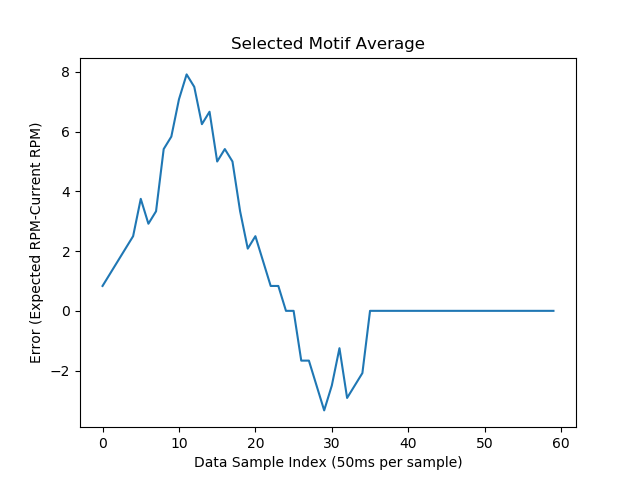}
    \captionof{figure}{Cancellation motif associated with Figure \ref{fig:tri1}}
    \label{fig:tri2}
\end{figure}

\noindent
In this particular signal denial process, it can be observed that the signal is incompletely cancelled, due to a sub-optimal "off-target" motif sample, in which only one portion of the true interference waveform is sampled. Exploring motif sampling and ranking is the next step of many in optimizing CEEDS. The way samples and rankings are handled can be modified to decrease the chance of an off-target motif sample as in Figure \ref{fig:tri2}. Methods such as increasing the number of motifs generated from the time series data and modifying thresholds and processes of the ranking algorithm could directly decrease the odds of an off-target motif. This process could also be accelerated with a preliminary filter on the candidate motifs by adding a minimum amplitude deviance, so as not to pick up on background stochastic noise rather than the signal in question.
\section{Conclusion}
CEEDS is highly effective in the analysis and cancellation of periodic error patterns in time series data, reducing error by up to 70\% in ideal conditions, which means that used correctly, CEEDS far exceeds the performance of comparable algorithms at a significantly lower computation costs. It should be noted that while CEEDS is currently incapable of correcting for multiple cyclical errors or errors with nonstandard periods, a system which is capable of correcting for such interference would be possible using a similar architecture. However, given that our goal with this research was to establish the groundwork for a family of new control systems, we assert that such improvements can be made in further research into the limits of the capabilities of these systems.
\newpage
\label{sec:others}
\section{Supplementary Figures}
\begin{figure}[h!]
    \centering
    \captionsetup{type=figure}
    \includegraphics[scale=0.6]{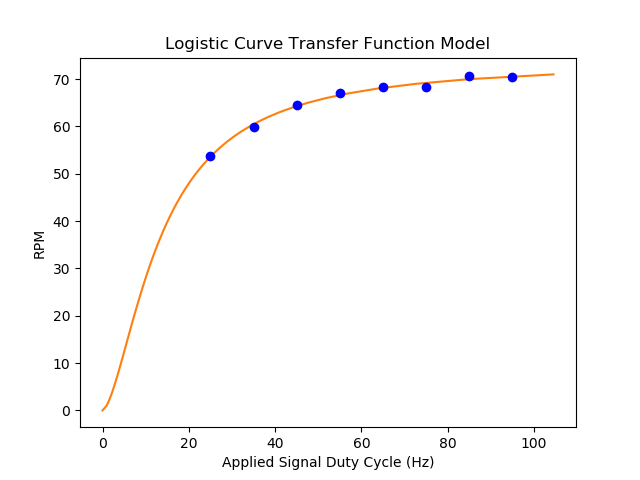}
    \captionof{figure}{Transfer function for PWM to RPM with data sample points from which the function was generated}
    \label{fig:tf1}
\end{figure}
\begin{figure}[h!]
    \centering
    \captionsetup{type=figure}
    \includegraphics[scale=0.6]{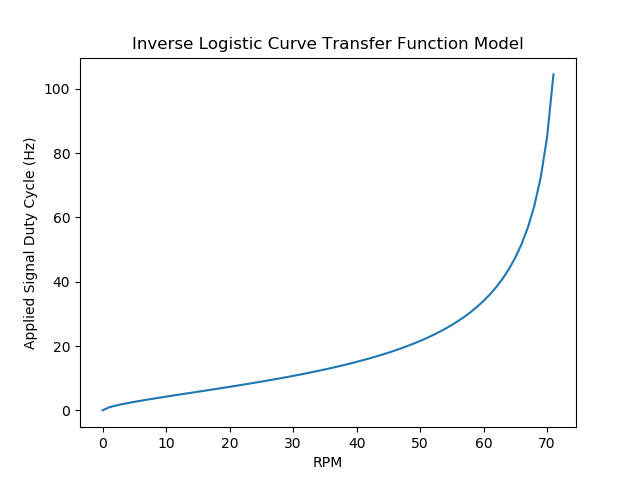}
    \captionof{figure}{Transfer function for RPM to PWM}
    \label{fig:tf2}
\end{figure}
\begin{figure}[h!]
    \centering
    \captionsetup{type=figure}
    \includegraphics[scale=0.6]{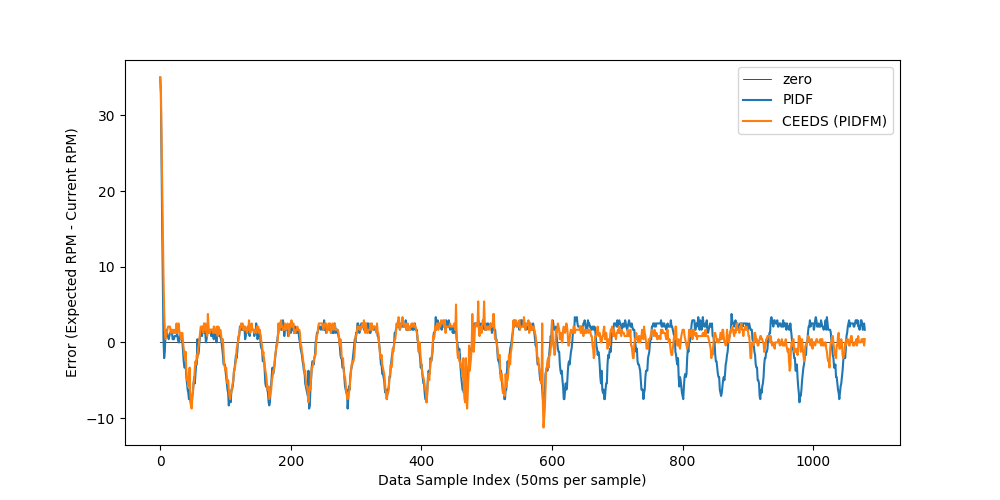}
    \captionsetup{justification=centering,margin=2cm}
    \captionof{figure}{Denial of triangular wave repeat trial. An error reduction from data sample 50 to the end of the data was calculated to be 31\%, and 66\% from data sample 600 to the end of the collected data.
}
    \label{fig:tri3}
\end{figure}

\begin{figure}[h!]
    \centering
    \captionsetup{type=figure}
    \includegraphics[scale=0.6]{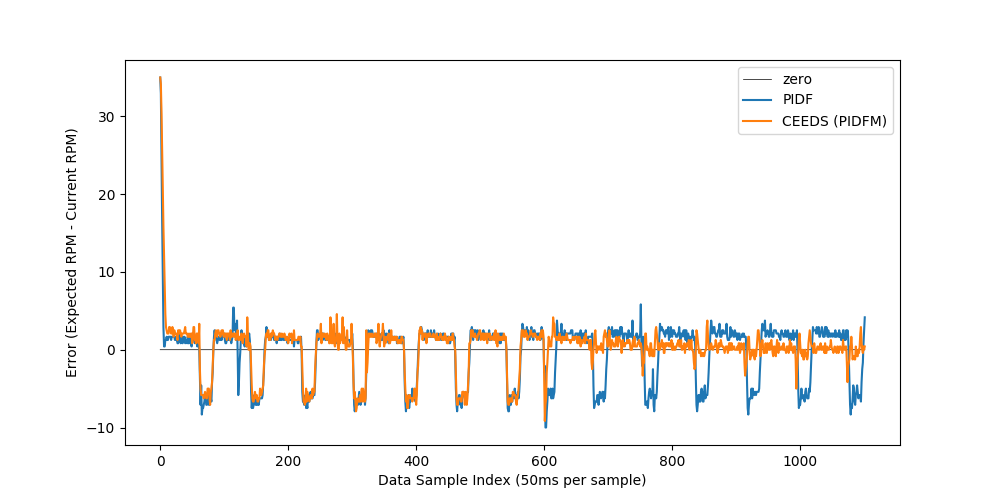}
    \captionsetup{justification=centering,margin=2cm}
    \captionof{figure}{Repeated square wave trial with less than ideal time series data due to stochastic interference. An error reduction from data sample 50 to the end of the data was calculated to be 37\%, and 72\% from data sample 600 to the end of the collected data.}
    \label{fig:square3}
\end{figure}

\begin{figure}[h!]
    \centering
    \captionsetup{type=figure}
    \captionsetup{justification=centering,margin=2cm}
    \includegraphics[scale=0.6]{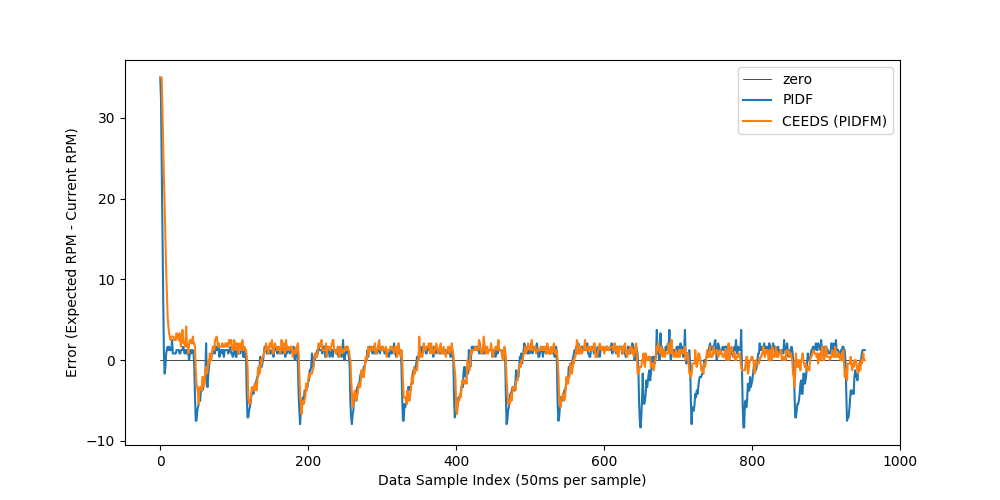}
    \captionof{figure}{Example of cancellation of sawtooth wave trial two. An error reduction from data sample 50 to the end of the data was calculated to be 21\%, and 56\% from data sample 600 to the end of the collected data. The decrease in efficacy during this trial can be attributed to an off-target motif sample being applied.}
    \label{fig:tri5}
\end{figure}

\begin{figure}[h!]
    \centering
    \captionsetup{type=figure}
    \includegraphics[scale=0.6]{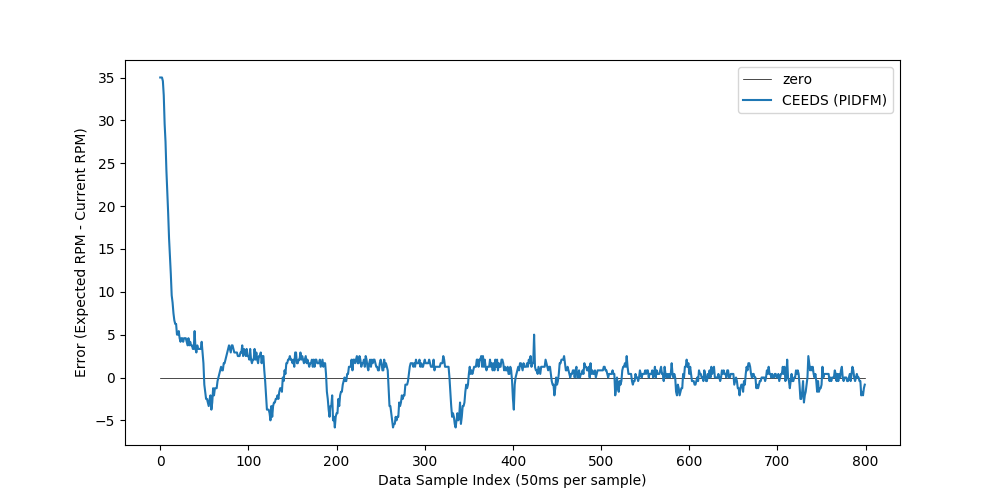}
    \captionsetup{justification=centering,margin=2cm}
    \captionof{figure}{Example of cancellation of a smaller triangular wave with an off-target motif alignment. No error reduction or non CEEDS equivalent data was generated for this trial.}
    \label{fig:tri4}
\end{figure}

\begin{figure}[h!]
    \centering
    \captionsetup{type=figure}
    \captionsetup{justification=centering,margin=2cm}
    \includegraphics[scale=0.6]{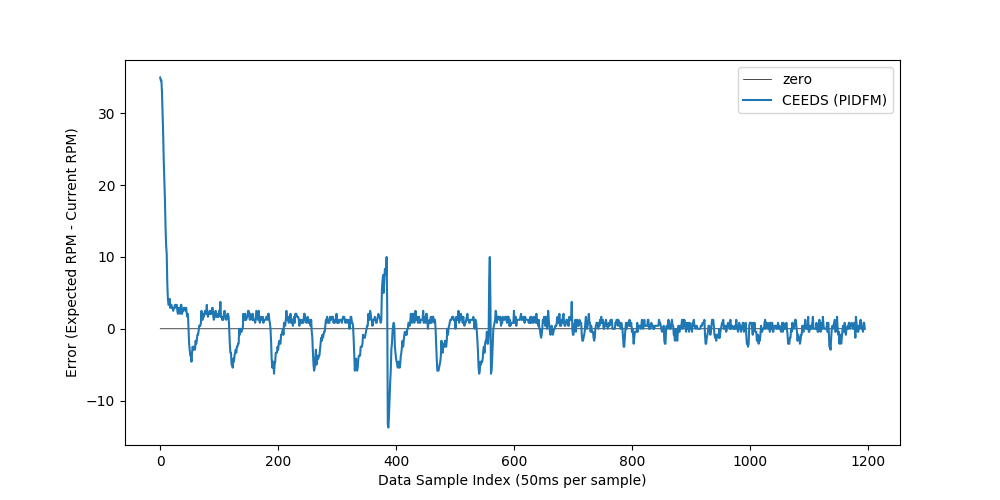}
    \captionof{figure}{Example of cancellation of sawtooth wave with less than ideal time series data, demonstrating CEEDS' ability to generate an accurate cancellation signal in lieu of "clean" sample data. Error spikes were unintentionally introduced, but the figure was included as a demonstration of algorithmic resilience.}
    \label{fig:tri6}
\end{figure}

\clearpage
\section{Acknowledgement}
We would like to thank and acknowledge Andrew Van Benschoten, Ph.D, and Tyler Marrs from the \href{https://matrixprofile.org}{Matrix Profile Foundation} for their guidance and technical assistance throughout the development of this research.

\bibliographystyle{IEEEtran}
\bibliography{references}{}

\begin{thebibliography}{10}
\providecommand{\url}[1]{#1}
\csname url@samestyle\endcsname
\providecommand{\newblock}{\relax}
\providecommand{\bibinfo}[2]{#2}
\providecommand{\BIBentrySTDinterwordspacing}{\spaceskip=0pt\relax}
\providecommand{\BIBentryALTinterwordstretchfactor}{4}
\providecommand{\BIBentryALTinterwordspacing}{\spaceskip=\fontdimen2\font plus
\BIBentryALTinterwordstretchfactor\fontdimen3\font minus
  \fontdimen4\font\relax}
\providecommand{\BIBforeignlanguage}[2]{{%
\expandafter\ifx\csname l@#1\endcsname\relax
\typeout{** WARNING: IEEEtran.bst: No hyphenation pattern has been}%
\typeout{** loaded for the language `#1'. Using the pattern for}%
\typeout{** the default language instead.}%
\else
\language=\csname l@#1\endcsname
\fi
#2}}
\providecommand{\BIBdecl}{\relax}
\BIBdecl

\bibitem{pidIntegrating}
\BIBentryALTinterwordspacing
H.~J. Kwak, S.~W. Sung, and I.-B. Lee, ``On-line process identification and
  autotuning for integrating processes,'' \emph{Industrial {\&} Engineering
  Chemistry Research}, vol.~36, no.~12, pp. 5329--5338, Dec 1997. [Online].
  Available: \url{https://doi.org/10.1021/ie9605600}
\BIBentrySTDinterwordspacing

\bibitem{LimitsPID1}
\BIBentryALTinterwordspacing
S.~W. Sung and I.-B. Lee, ``Limitations and countermeasures of pid
  controllers,'' \emph{Industrial {\&} Engineering Chemistry Research},
  vol.~35, no.~8, pp. 2596--2610, Jan 1996. [Online]. Available:
  \url{https://doi.org/10.1021/ie960090+}
\BIBentrySTDinterwordspacing

\bibitem{LimitsPID2}
D.~P. {Atherton} and S.~{Majhi}, ``Limitations of pid controllers,'' in
  \emph{Proceedings of the 1999 American Control Conference (Cat. No.
  99CH36251)}, vol.~6, 1999, pp. 3843--3847 vol.6.

\bibitem{nnPID}
\BIBentryALTinterwordspacing
J.~Kang, W.~Meng, A.~Abraham, and H.~Liu, ``An adaptive pid neural network for
  complex nonlinear system control,'' \emph{Neurocomputing}, vol. 135, pp. 79
  -- 85, 2014. [Online]. Available:
  \url{http://www.sciencedirect.com/science/article/pii/S092523121301134X}
\BIBentrySTDinterwordspacing

\bibitem{nnStability}
K.~Tanaka, ``An approach to stability criteria of neural-network control
  systems,'' \emph{IEEE Transactions on Neural Networks}, vol.~7, no.~3, pp.
  629--642, 1996.

\bibitem{nnSUCKS}
\BIBentryALTinterwordspacing
T.~J. [de Vries], W.~J. Velthuis, and J.~[van Amerongen], ``Learning
  feed-forward control: A survey and historical note,'' \emph{IFAC Proceedings
  Volumes}, vol.~33, no.~26, pp. 881 -- 886, 2000, iFAC Conference on
  Mechatronic Systems, Darmstadt, Germany, 18-20 September 2000. [Online].
  Available:
  \url{http://www.sciencedirect.com/science/article/pii/S147466701739256X}
\BIBentrySTDinterwordspacing

\bibitem{telescoper}
E.~D. {Klenske}, M.~N. {Zeilinger}, B.~{Schölkopf}, and P.~{Hennig},
  ``Gaussian process-based predictive control for periodic error correction,''
  \emph{IEEE Transactions on Control Systems Technology}, vol.~24, no.~1, pp.
  110--121, 2016.

\bibitem{mpSUPERHUMAN}
S.~{Gharghabi}, Y.~{Ding}, C.~M. {Yeh}, K.~{Kamgar}, L.~{Ulanova}, and
  E.~{Keogh}, ``Matrix profile viii: Domain agnostic online semantic
  segmentation at superhuman performance levels,'' in \emph{2017 IEEE
  International Conference on Data Mining (ICDM)}, 2017, pp. 117--126.

\bibitem{mpMOTIFS}
\BIBentryALTinterwordspacing
C.-C.~M. Yeh, Y.~Zhu, L.~Ulanova, N.~Begum, Y.~Ding, H.~A. Dau, Z.~Zimmerman,
  D.~F. Silva, A.~Mueen, and E.~Keogh, ``Time series joins, motifs, discords
  and shapelets: a unifying view that exploits the matrix profile,'' \emph{Data
  Mining and Knowledge Discovery}, vol.~32, no.~1, pp. 83--123, Jan 2018.
  [Online]. Available: \url{https://doi.org/10.1007/s10618-017-0519-9}
\BIBentrySTDinterwordspacing

\bibitem{VanBenschoten2020}
\BIBentryALTinterwordspacing
A.~V. Benschoten, A.~Ouyang, F.~Bischoff, and T.~Marrs, ``Mpa: a novel
  cross-language api for time series analysis,'' \emph{Journal of Open Source
  Software}, vol.~5, no.~49, p. 2179, 2020. [Online]. Available:
  \url{https://doi.org/10.21105/joss.02179}
\BIBentrySTDinterwordspacing

\bibitem{algorithms}
\BIBentryALTinterwordspacing
{Matrix Profile Foundation}. (2020) {Algorithms – Matrix Profile}. [Online].
  Available: \url{https://matrixprofile.docs.matrixprofile.org/Algorithms.html}
\BIBentrySTDinterwordspacing

\end{thebibliography}

\end{document}